# Predicting mental health using social media: A roadmap for future development

**Ramin Safa***
Department of Computer Engineering,
Ayandegan Institute of Higher Education,
Tonekabon, Iran

**S. A. Edalatpanah**
Department of Applied Mathematics,
Ayandegan Institute of Higher Education,
Tonekabon, Iran

**Ali Sorourkhah**
Department of Management, Ayandegan
Institute of Higher Education,
Tonekabon, Iran

## Abstract

Mental disorders such as depression and suicidal ideation are hazardous, affecting more than 300 million people over the world. However, on social media, mental disorder symptoms can be observed, and automated approaches are increasingly capable of detecting them. The considerable number of social media users and the tremendous quantity of user-generated data on social platforms provide a unique opportunity for researchers to distinguish patterns that correlate with mental status. This research offers a roadmap for analysis, where mental state detection can be based on machine learning techniques. We describe the common approaches for predicting and identifying the disorder using user-generated content. This research is organized according to the data collection, feature extraction, and prediction algorithms. Furthermore, we review several recent studies conducted to explore different features of candidate profiles and their analytical methods. Following, we debate various aspects of the development of experimental auto-detection frameworks for identifying users who suffer from disorders, and we conclude with a discussion of future trends. The introduced methods can help complement screening procedures, identify at-risk people through social media monitoring on a large scale, and make disorders easier to treat in the future.

*Corresponding author: safa@aihe.ac.ir

# 1. Introduction

WHO[1] explains mental disorders as various issues with a broad range of symptoms. A general characteristic of these disorders is the presence of abnormal behavior, thoughts, and emotions. The following disorders are a few examples: Depression, anxiety, bipolar, eating, Post-Traumatic Stress, schizophrenia, problems associated with drug abuse, and intellectual disabilities. Approximately 970 million people worldwide suffer from mental disorders, according to the Mental Disorders Fact Sheet[2] published by the WHO in June 2022. Since the beginning of the COVID-19 pandemic in 2020, many people suffering from anxiety and depression have increased significantly. Approximately 83 million European Economic Area (EEA) citizens (between 18-65) have experienced one or more mental disorders over the past year (Ríssola et al., 2021). In addition, without appropriate treatment, someone can experience psychotic episodes, disability, self-harm attitudes, or commit suicide. Symptoms of these mental disturbances must therefore be recognized at early stages to avoid each unwanted consequence.

As reported by the Lancet Commission on global mental health and sustainable development in 2018, mental health disorders contribute to an increasing number of diseases worldwide. While social services often fail to meet the same standards as physical health services. (Patel et al., 2018). Depressive disorders were ranked the most prevalent mental health problem worldwide in the 2013 GBD[3] study, continued by bipolar, schizophrenia, and anxiety disorder. According to the GBD 2019[4] reports, 1.2% (more than 815,000 cases) of the United Kingdom's population suffer from bipolar disorder (Harvey et al., 2022).

Social networks offer unique opportunities to study interpersonal relationships and the social context of modern societies, particularly among the under-25s, their primary consumers (Bersani et al., 2022). As, among youth, suicide is a primer ground of death, this examination is statistically crucial (Organization, 2021), even though suicide rates generally increase with age. In addition, the number of suicide attempts among young people is also exceptionally high. In adolescence, extensive brain changes will shape cognitive development forced by social media (Crone & Konijn, 2018). It is also possible to reduce intervention costs using social workers in mental health services (Rice et al., 2016).

Social media platforms have become increasingly widespread among people nowadays to express their feelings and moods. As a result of this phenomenon, researchers and healthcare professionals can classify linguistic indicators related to mental illnesses like schizophrenia, depression, and suicide. By examining an individual's language use, we can gain valuable insight into their mental state, personality, and social and emotional condition (Uban et al., 2021). Studies have shown that using language attributes to describe one's mental state, nature, and even personal values can be a strong indicator of their current mental state, character, and values. The interaction of language and clinical disorders has been investigated in several studies. According to conclusions, "speech content can provide a unique window into thoughts" (Ríssola et al., 2021), making it possible to directly diagnose mental disorders, for instance, resulting from addiction. As people increasingly turn to platforms such as Twitter, Facebook, and Reddit to express their opinions, feelings, and moods, tools may be developed to diagnose various mental health issues (Guntuku et al., 2017; Thorstad & Wolff, 2019). The language and emotions used in social media posts can illuminate feelings such as worthlessness, guilt, and helplessness. Thus the symptoms of psychological disorders can be characterized this way. Coppersmith et al. (Coppersmith, Dredze, Harman, & Hollingshead, 2015) argued that these kinds of concerns are often disclosed on social media for a variety of reasons, including pursuing or providing encouragement, changing society's stigma or taboo against mental illness, or explaining certain behaviors (Ríssola et al., 2021).

---

[1] World Health Organization
[2] https://who.int/news-room/fact-sheets/detail/mental-disorders
[3] Global Burden of Disease
[4] https://healthdata.org/gbd/2019

Researchers have been studying the possibility of detecting mental state alterations through language for several years. As a result of social data, health specialists can identify people and communities at risk. For example, experts can identify individuals who need immediate care through a large-scale monitoring program. As the subject has received considerable attention lately, it has become necessary to organize and summarize conventional approaches and the latest trends in a scoping review. This chapter outlines a roadmap and current advances for evaluating mental states and identifying disorders based on user digital footprints on social platforms. The assessment approaches and feature extraction process were also categorized. Thus, young researchers can understand the latest trends and developments in the area.

Below is a summary of the remainder of the chapter. Following the literature on mental disorders, social platforms, and social data analysis, Section 2 discusses these concepts. This chapter demonstrates how social data, information retrieval, NLP[5], and machine learning techniques are used to build mental health assessment tools and develop DSSs[6] in psychiatry. Section 3 will present various data collection methods and evaluations in the field of research. Feature engineering and preprocessing big social data are discussed in Section 4. Prediction methods and evaluation methods are also discussed in sections 5 and 6. In section 7, there are concluding remarks including challenges and possible directions for future research. Finally, Section 8 summarizes the findings and presents future directions.

## 2. Mental disorders and big social data

Based on WHO, mental health involves understanding yourself, coping with everyday stress, working productively, and positively contributing to society. According to estimates, the global economy loses trillions of dollars to mental disorders (Whiteford et al., 2016). Some of the most known disorders are summarized in Table 1. There is, however, a significant decline in treatment and quality of care for those with mental illnesses because of resource shortages (Docrat et al., 2019; Lee et al., 2022; Petersen et al., 2019). It is also declared that many countries are suffering from a lack of psychiatrists (Hanna et al., 2018). Furthermore, mental health professionals have insufficient tools and methods for decision-making on care-related

---

[5] Natural Language Processing
[6] Decision Support System

issues, such as accurate diagnoses (Kilbourne et al., 2018).

Recent pandemics have also worsened the global mental health crisis (Johnson et al., 2021; Lee et al., 2022). Therefore, governments, policymakers, and mental health professionals (like psychiatrists and counselors) require innovative tools to help them in diagnostic decisions made with greater efficiency and accuracy (Thieme et al., 2020; Tutun et al., 2022).

**Table 1.** Common mental disorders definition

| Mental Disorder | Definition |
|---|---|
| Depression | There is a difference between depression and mood swings or short-lived emotional reactions to daily experiments; A mental state causing painful symptoms adversely disrupts normal activities (e.g., sleeping). For at least two weeks, the person experiences depressive moods (sad, irritable, empty) or a lack of interest in activities for most of the day or the week. |
| Anxiety | Several behavioral disturbances are associated with anxiety disorders, including excessive fear and worry. Severe symptoms cause significant impairment in functioning cause considerable distress. Anxiety disorders come in many forms, such as social anxiety, generalized anxiety, panic, etc. |
| Bipolar Disorder | An alternating pattern of depression and manic symptoms is associated with bipolar disorder. An individual experiencing a depressive episode may feel sad, irritable, empty, or lose interest in daily activities. Emotions of euphoria or irritability, excessive energy, and increased talkativeness can all be signs of manic depression. Increased self-esteem, decreased sleep need, disorientation, and reckless behavior may also be signs of manic depression. |
| Post-Traumatic Stress Disorder (PTSD) | In PTSD, persistent mental and emotional stress can occur after an injury or severe psychological shock, characterized by sleep disturbances, constant vivid memories, and dulled response to others and the outside world. People who re-experience symptoms may have difficulties with their everyday routines and experience significant impairment in their performance. |
| Seasonal Affective Disorder (SAD) | In most cases, SAD occurs in the fall/winter and enters remission in the spring/summer, although in some cases, it may happen in the summer and remit in the autumn and winter. A majority of the cause-and-effect mechanisms of SAD have not yet been discovered. However, several hypotheses have been posed regarding the disease, and they promise to deliver new information to scientists. |
| Schizophrenia | A schizophrenic disorder is characterized by episodes of psychosis that occur continuously or recur continuously. Disorganized thinking, hallucinations, and delusions are some of the significant symptoms. |

|  | Other symptoms are apathy, social withdrawal, and a decreased expression of emotions. Most symptoms develop gradually, begin during young adulthood, and do not resolve in most cases. |
| --- | --- |
| Eating Disorders | In eating disorders, there is a persistent disturbance in eating behavior, along with distressing thoughts and feelings. These conditions negatively impact physical, psychological, and social functioning. Disorders like anxiety and obsessive-compulsive disorders often coexist with eating disorders. |

Even with various diagnostic guidelines and tools available, diagnosing patients accurately and efficiently remains challenging. As a result, recent literature suggests that more research should be done on how cutting-edge technologies (such as data science) can be used to create clinical DSSs assisting psychologists and directing health informatics developers (Balcombe & De Leo, 2021). According to Figure 1, social media-based e-mental health research has the following conceptual framework.

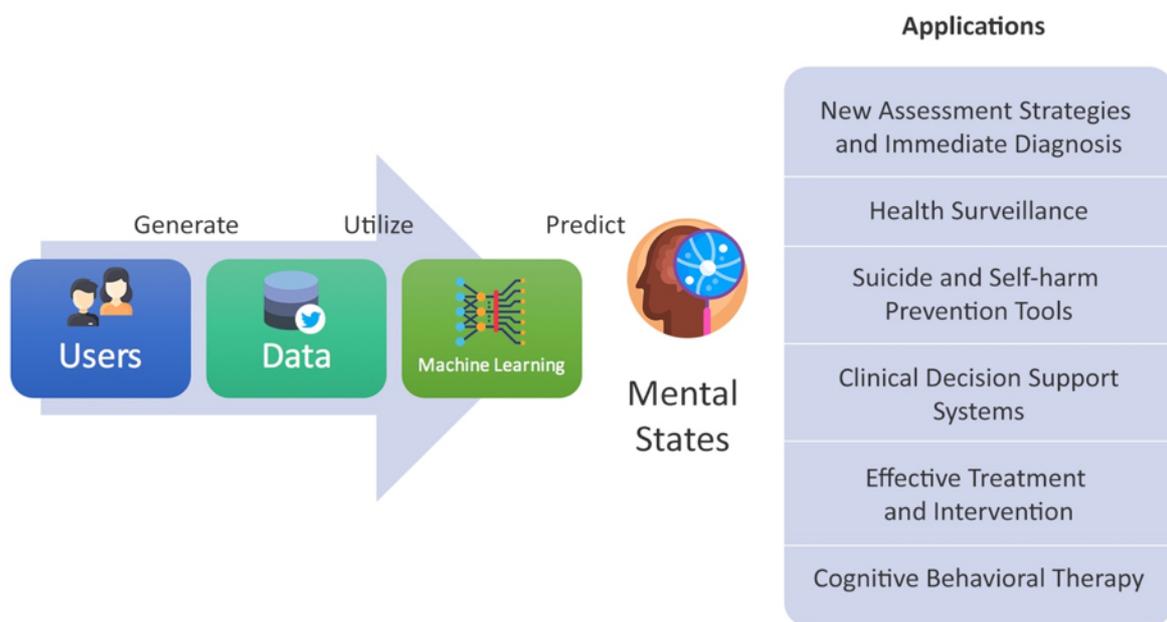

**Figure 1.** Research on e-mental health through social media: a conceptual framework

This chapter precisely responds to social media-based e-mental health assessment and addresses diagnosing mental disorder challenges concerning the guidelines and tools used. We aim to show a road map to build a social media-based e-mental health assessment tool and present the requirements to develop an artificial intelligence-based DSS.

The development of data science has been driven by increasing data volumes, variety, analytics, visualization, and computational power. Using information retrieval and machine learning techniques, researchers are extracting information from massive amounts of data and using them to expound and construct classification models in numerous fields. Researchers have detected meaningful patterns in healthcare datasets using data science methods. Section 3 discusses how the data of social media mining are applied to detect mental disorder symptoms.

Social media is social by nature. As a result, raw data may readily reveal and quantify social interaction patterns, which are crucial to mental state assessments. Because social media is (semi) anonymous and open, people are more likely to socialize and share their information (De Choudhury, Counts, et al., 2013a). Sharing content concerning your everyday life and announcing prominent life landmarks on social media is expected. The way the subject expresses herself can provide a considerable amount of information about their mental state and emotional conditions, as introduced in the introduction. Thus, a user's social data content can convey feelings such as helplessness, worthlessness, or guilt due to the language employed and the emotions conveyed. Different mental disorders can thus be identified and characterized in this way. Data collected from social platforms occurs non-reactive and can complement conventional data in a valuable way.

With billions of users, Facebook is the leading social media network, followed by Twitter and Instagram. Many features are available on Facebook, including creating a profile, uploading files, sending messages, and being in contact. Users can post information about themself, whether about their occupation, religion, political views, or favorite movies and musicians. With Twitter, users can tweet, short stream up to 280 characters, and follow other individuals' updates. Young adults between 18 and 29 tend to be more attracted to it (Lopez-Castroman et al., 2020). Based on Statista[7], In 2019, Twitter had 290 million active users (monthly), and the population is anticipated to be over 340 million by 2024. Users can upload photos and videos on Instagram while exposing their current location. Reddit is also another social network that can be analyzed for a variety of reasons. In recent years, these platforms have become increasingly popular for expressing opinions, interacting with others, and sharing feelings. Hootsuite's well-established social media management platform claims that social media users have recently grown by more than two hundred million. Globally, social media usage has increased by more than 5%, reaching 59% of the population. In addition, unique users have risen by some 520 million, representing an annual growth rate of over 13%. Figure 2 illustrates the growing trend of social media users in the last eleven years.

---

[7] https://statista.com/statistics/303681/twitter-users-worldwide/

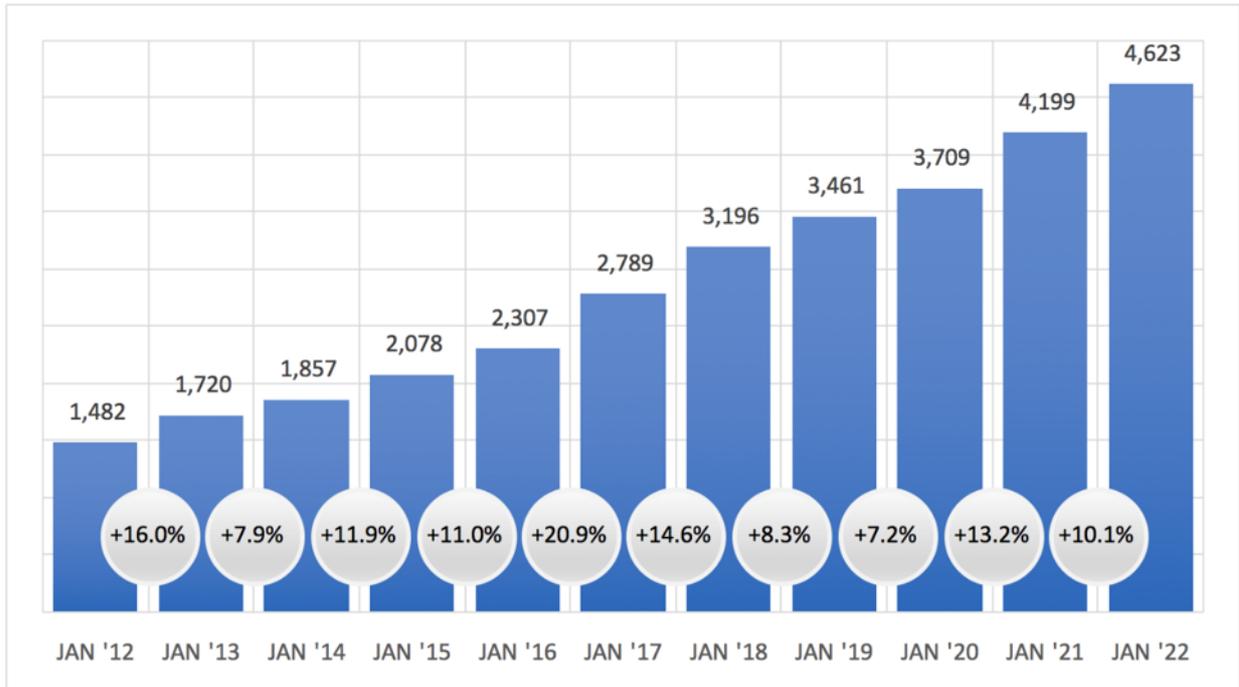

**Figure 2.** Users of social media over time (in millions)

It results in social big data, which contains worthwhile information about people's behavior, moods, and interests (Javadi et al., 2020; Kang et al., 2016; Martínez-Castaño et al., 2020), which covers a considerable number of fields containing machine learning, data and graph mining, statistics, information retrieval, linguistics, NLP, and text mining. In addition to health and e-commerce, their applications can be extended to many other areas (Bello-Orgaz et al., 2016; Kumar et al., 2020).

In a study of a large set of tweets, self-reported signs were found to be the most trustworthy signal for predicting syndrome outbreaks (Krieck et al., 2011). Similarly, studies have found that social networks can detect trends in disease outbreaks, such as flu spread investigations (Sooknanan & Mays, 2021) or depression (Lopez-Castroman et al., 2020). Using this data, Researchers can deeply understand the users' behavior.

Monitoring population and mental health are increasingly being conducted through social media (Conway & O'Connor, 2016). Online screening tools, like medical DSSs, are practical and may serve as more common assessment strategies in the future (Ebert et al., 2019); likewise, health surveillance tools detect mental disorders signs, despite inefficient traditional approaches relying heavily on interviews and surveys. Distinguishing recognized signals from user-generated social media content might shape new mental disorders screening forms. A potential advantage of automated social data analysis is the ability to detect early warning signs. According to previous studies, machine learning may identify depression early using language patterns as indicators (Coppersmith et al., 2018; Loveys et al., 2017; Martínez-Castaño et al., 2020; Plaza-del-Arco et al., 2020).

## 3. Assessment strategies

By using an automated process, it would be possible to identify the signs of an individual's disorder in their behavior and target them for a more thorough assessment. There has been an increase in the number of researchers investigating mental health within the context of social media in recent years, examining the association between social media use and behavioral patterns and disorders such as anxiety, depression, suicidality, and stress. This section discusses methods to predict mental disorders and five commonly used approaches. A discussion of the differences is then provided, along with directions for future research.

To collect social data with associated information about people's mental health, several approaches have been studied. Figure 3 shows how data is gathered from sources by either recruiting participants to take a survey and/or sharing their social network profile information. Users' tweets can be searched for specific keywords to detect (and collect all tweets from) those who have shared diagnoses, user language on tweets, or mental illness-related forums mentioning mental disease keywords can be collected. Using public data can be more cost-effective and faster than administering surveys if a much larger sample is collected. However, clinical information and survey-based assessments generally have a higher degree of validity (Guntuku et al., 2017). A more detailed examination of these methods will follow in the next sections.

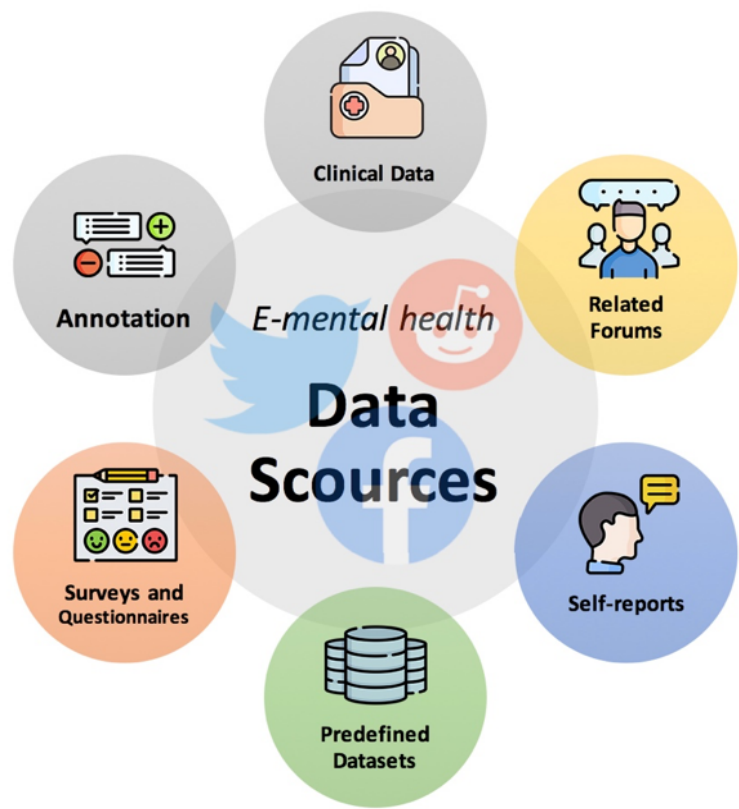

**Figure 3.** The data source for e-mental health research

### 3.1. Questionnaires and surveys

The psychological characteristics of users are identified through questionnaires and surveys, which are highly used in the social and behavioral sciences (Spiro, 2016). Using psychometric self-report surveys to assess mental illness is highly reliable and valid. The use of self-report surveys in psychology and epidemiology is second only to clinical interviews (Neter et al., 1996). CES-D[8], PHQ-9[9], and BDI[10] are the most common feedback form to assess participants' depression levels. Among the relevant instruments for detecting suicidal ideation, and measuring well-being, are the SPS[11] and SWLS[12] (Safa et al., 2022). Participants could participate in the research by answering questionnaires and following data collection guidelines by visiting data donation websites or crowdsourcing platforms, like OurDataHelps[13] or MTurk[14] (Braithwaite et al., 2016b; Coppersmith et al., 2018; Schwartz et al., 2016).

Predictive models in this emerging literature can be most accurately built using survey responses. As a result of the high costs associated with this method, more publicly accessible assessment criteria are being used, such as those outlined in the following sections. Combining social data with surveys is additionally not without challenges in practice. Social media identities can be asked for in surveys or questionnaires as part of participants' contact information. Whether or not participants volunteer this information is up to them (Spiro, 2016).

### 3.2. Self-declared mental health status

Public data is used in a number of studies. Social media APIs[15] are used to collect related posts using keywords/phrases or regular expressions. Because there were no standards for data collection, there was a need for a custom data capture mechanism for certain data sources. As an example, custom tools or web apps developed to connect to the Facebook APIs were used to gather datasets for Facebook-based experiments (De Choudhury et al., 2014; Park et al., 2015). Likewise, several studies have explored mental disorder cues using Twitter APIs (Braithwaite et al., 2016a; Safa et al., 2022; Wang et al., 2017). There have been similar approaches taken for Reddit (Boettcher, 2021), Sina Weibo (Lv et al., 2015), and Instagram APIs (Ferwerda & Tkalcic, 2018).

The leading queries for post-retrieval are "suicide," "self-harm," "kill yourself," "want to die," and "I was diagnosed with [disorder]" (Coppersmith et al., 2014; Losada & Crestani, 2016; Ríssola et al., 2020). An example of publicly available data is self-declared mental illness diagnoses on Twitter. Self-reported diagnosis is based on regular expressions of this type. API extraction leads to a set of tweets that must be assessed before analysis. Note that posts negate suicide ideation, discuss the suicide of others, or report news of suicide removed from the collection. In the same way, posts without hypothetical statements, negations, or quotations are selected as positive samples in the case of self-report diagnosis.

---

[8] Center for Epidemiologic Studies Depression Scale
[9] Patient Health Questionnaire
[10] Beck Depression Inventory
[11] Suicide Probability Scale
[12] Satisfaction with Life Scale
[13] https://ourdatahelps.org
[14] https://mturk.com
[15] Application Programming Interface

Twitter users' self-declared diagnoses were used to predict whether they had PTSD or depression to facilitate collaboration between computer scientists and clinical psychologists in the 2015 CLPsych[16] workshop (Coppersmith, Dredze, Harman, Hollingshead, et al., 2015). During the study, participating teams constructed language topic models (Resnik et al., 2015), identified words most related to disorder status, considered character sequences as features, and built relative counts of N-grams present in all disorders statuses using a rule-based approach (Pedersen, 2015). Prediction performance was highest for the latter. Researchers have been interested in this method, which has recently been used to develop automatic tools for detecting depression (Safa et al., 2022).

### 3.3. Forum membership

Another source of available mental health information is online forums and discussion websites. Generally, they enable people to discuss stigmatized mental health issues in an open space, receive and provide emotional support, and ask for advice. This approach commonly uses Reddit and Facebook for data collection.

De Choudhury et al. (De Choudhury et al. 2016) looked at posts of Reddit users (via subreddits) who addressed mental health issues and then spoke about suicidal ideation in the future. It was discovered that these shifts were characterized by poor linguistic coordination, reduced social engagement, heightened self-attentional focus, and a sense of hopelessness, impulsiveness, anxiety, and loneliness in shared content. A study by Tadesse et al. (Tadesse et al., 2019) looked for ways to detect depression on Reddit social media and solutions for detecting depression through effective performance increases. Language usage and depression were found to be closely related using NLP and text classification techniques. In summary, depressing accounts were more likely to contain language predictors of depression, with an emphasis on the present and the future, referring to feelings of sadness, anger, anxiety, or suicidal thoughts.

### 3.4. Posts annotation

Manually reviewing and annotating posts that contain mental health keywords is the next source of publicly-available data. The language of social media posts can predict annotations. Social media posts are coded concerning pre-established categories by annotations (Kern et al., 2016). Annotation studies on depression typically look for posts where users discuss their own experiences with depression (Cavazos-Rehg et al., 2016). In addition to guidelines on recognizing depression symptoms, annotators are provided with a reduced set of symptoms, including disturbed sleep, depressed mood, and fatigue, as described in clinical assessment manuals such as the DSM-5[17] (Association, 2013). Furthermore, annotations have been applied to distinguish between stigmatizing and insulting mentions of mental illness from expressing, sharing, or supporting helpful information with those with mental disorders. The annotations of posts are generally used as a supplementary method for revealing life conditions accompanying mental illness (such as education, employment, housing, or weather problems) not apprehended by conventional depression diagnostic indicators (Mowery et al., 2015).

---

[16] https://clpsych.org
[17] Diagnostic and Statistical Manual of Mental Disorders

An innovative dataset for CAMS[18] has been developed by Garg et al. (Garg et al., 2022); Using two separate datasets: crawling and annotating 3155 Reddit posts and re-annotating 1896 instances from the available SDCNL dataset[19] (Haque et al., 2021) for interpretable causal analysis, the authors present a causal analysis annotation schema. Their experimental results showed that a classic Logistic Regression model performed better than a CNN-LSTM[20] model on the CAMS dataset.

### 3.5. Other methods
To provide access to individuals with mental illnesses, computer scientists should collaborate with physicians and psychologists. Predictive models could be built based on real patients' data. It is possible to create very reliable documents based on clinical reports along with social data. Social network platforms can be analyzed using user-generated content and behavior to discover meaningful patterns by having the mental state of individuals and the appropriate permissions.

aAnother option is to use predefined datasets, in which data are collected by other researchers and pass the initial evaluation phase to identify class labels. The myPersonality[21] project, CLPsych (Losada & Crestani, 2016), and eRisk[22] workshops (Losada et al., 2020) provide users' social data and psychometric test scores for academic purposes, are three well-known datasets in the field in question. AutoDep[23] is another new dataset that is provided automatically and used to examine more features in the social data produced by Twitter users with symptoms of depression (Safa et al., 2022).

### 4. Social Data Configuration
Building predictive models using extracted data is the process of automating the analysis of social networks. The social data collected on social media networks come from users' online activity. The content can include text, images, videos, context information (such as location tags), user biographical information, connections, and interests (Figure 4).

Numerous research applies textual content and linguistic patterns to recognize the most significant consequence of mental disorder prediction. Posting frequencies, hashtags, and times of posts are among users' common features. LIWC[24] (Pennebaker et al., 2015) analysis obtained some of these results. In related studies, the LIWC is a notable text analysis application for finding linguistic patterns (Chen, Sykora, Jackson, Elayan, et al., 2018; Chen, Sykora, Jackson, & Elayan, 2018; Loveys et al., 2017; Ríssola et al., 2020). Various psychologically meaningful categories are covered by dictionaries that psychologists manually construct. In addition to extracting positive or negative sentiments and personal pronouns from the textual content, it can extract potential signs from the text. As popular sentiment analysis tools, OpinionFinder (Wilson et al., 2005) and SentiStrength (Thelwall et al., 2010) were frequently used in selected studies to quantify textual sentiment (Bollen et al., 2011; Durahim & Coşkun, 2015). In many attempts to disclose latent topics from user posts, topic modeling techniques such as LDA[25] (Blei et al., 2003) have also been used as part of the content analysis (Ji et al., 2018; Paul & Dredze, 2011).

---

[18] Causal Analysis of Mental Health Issues in Social Media Posts
[19] https://ayaanzhaque.github.io/SDCNL
[20] Convolutional Long Short-Term Memory Neural Networks
[21] https://sites.google.com/michalkosinski.com/mypersonality
[22] https://erisk.irlab.org
[23] https://github.com/rsafa/autodep
[24] Linguistic Inquiry and Word Count

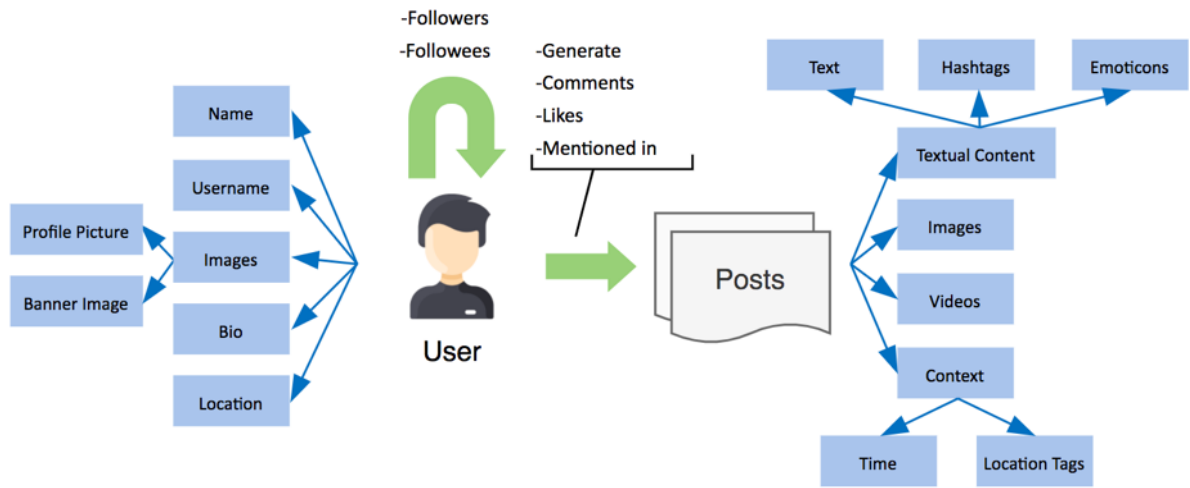

**Figure 4.** Principal components and relationships in social data

N-gram language models are also tools used to determine the probability of the occurrence of certain characters and word sequences. Traditional word-based approaches are not ideal in social media texts because of shortening and spelling errors (Chua et al., 2019). Hence, to extract features, we can employ character/word n-grams and use the tf-idf technique (Safa et al., 2022). After the features have been extracted, they are preserved as independent variables in a model, such as linear regression (Neter et al., 1996) or SVM[26] (Cortes & Vapnik, 1995), to signify the dependent variable (mental disorder). A machine learning technique is utilized for training a predictive model using training data and evaluated by the test data to avoid overfitting (cross-validation). As one of the different metrics, the prediction performances, which we will discuss in the evaluation section, are presented.

An example of Twitter data analysis using surveys involves collecting information on social network accounts while administering the survey to a well-defined study population. It is possible to sample social data systematically using Twitter account identifiers (User_IDs). According to Figure 4, we can access various observational data types using the Twitter API. Each class must be queried separately. The most recent tweets of a user are available at the time of querying (limited to 3,200 recent tweets). Typical data collection can conduct at regular intervals to resume longitudinal observation. To reach the target dataset, several standard steps must be done with the users' data. The first step was to clean and preprocess the data to ensure they were suitable for the analysis algorithms. Feature engineering selects a subset of features in the next step. In order to fit the model, the desired characteristics and information extracted from the questionnaires are analyzed. Following the testing phase, an evaluation is conducted based on the test set (Safa et al., 2022). In the following, we will take a quick look at these steps.

---

[25] Latent Dirichlet Allocation
[26] Support Vector Machines

### 4.1. Preprocessing

It is typical for the corpus of data to be preprocessed to remove unsuitable samples and to clean and prepare the data for analysis. Participants' questionnaires and pieces of information, usually extracted from studies, can contain unworkable and vague details to enhance prediction and classification accuracy. The dataset excludes individuals whose profiles lack sufficient information. A number of people with low activity levels were removed since they published fewer than a defined number of posts. The questionnaires excluded individuals whose completion time was abnormally short or long (Wongkoblap, 2020). Those whose responses to two different questionnaires were not correlated were also excluded. Each post was checked for written language during the data cleaning (Safa et al., 2022). As a result, the available tools were suitable for analyzing the posts.

In the preprocessing step, stop-words are usually removed, sentences are segmented, retweets, duplicates, URLs, special characters, mentioned usernames, and hashtags are handled, and lowercasing is done (Chen, Sykora, Jackson, Elayan, et al., 2018; Du et al., 2018; Ma et al., 2017). Additionally, emojis were converted to ASCII to ensure that machines could read the data. For ethical reasons, any potentially identifying usernames were also anonymized. The user may upload images and videos in a different format on some social networks, so it is essential to consider format conversion for visual content.

### 4.2. Feature engineering

Social network users may show signs of mental health problems based on a number of different features that can be extracted from their social networking profiles. Several studies have examined textual content to determine which factors are associated with mental health. Alternative research techniques, such as image analysis and social interaction analysis, have been used in other research projects.

Sentiment analysis is a popular tool in NLP and text mining, which classifies a given text's polarity into positive, negative, and neutral categories to understand the emotional expression (Kiritchenko et al., 2014). Several studies used LIWC (Pennebaker et al., 2015) to draw out mental disorders' signals from textual data (e.g., the occurrence of using pronouns "I" or "me" as determinants). OpinionFinder (Wilson et al., 2005) was used by Bollen et al. (Bollen et al., 2011), and SentiStrength (Thelwall et al., 2010) was used by Kang et al. (Kang et al., 2016) to carry out sentiment analysis. The literature likewise used VADER to exploit the benefits of rule-based modeling and lexicon-based characteristics for social media messages (Safa et al., 2022). Moreover, topic modeling was employed in many studies (Margus et al., 2021;

Schwartz et al., 2016) to extract topics from social content.

It is common for social media posts to contain a variety of emoticons. Consequently, some studies (Kang et al., 2016) looked into their use's meaning and mood states. Users of social network platforms can also post visual content in addition to text messages; Researchers have examined this data for mental disorder cues in some studies (Safa et al. 2022).

However, most of these studies rely heavily on textual content and only a few image-processing techniques (Chiu et al., 2020; Kumar & Garg, 2019). Kang et al. (Kang et al., 2016) used color composition and SIFT descriptors to discover emotional meaning from Twitter images. Reece et al. (Reece & Danforth, 2017) predict depression in Instagram users based on the image's hue, saturation, and brightness. Sharath et al. (Guntuku et al., 2019) have demonstrated that utilizing VGG-Net (Simonyan & Zisserman, 2015) image classifier together with image features like facial, aesthetics, color, and content are used to predict depression. More recently, Safa et al. (Safa et al. 2022) used user profile photos and header image content to discover depression's latent patterns. They used Imagga[27] tagging API to represent image content and generate a Bag-of-Visual-Words (BoVW) as a part of the analysis.

Users on social network platforms interact and form relationships with each other millions of times each day. A graph structure containing information about friendships, relationships, and interactions was analyzed to determine how mental disorders can be detected (e.g., assortative mixing patterns and interactions among depressed users) (Wang et al., 2017). It should also be noted that incorporating hashtags and context information (e.g., post time) can improve prediction accuracy, which demonstrated a difference in a study on the biggest social media platforms in China (Sina Weibo) (Mao et al., 2018)).

Following preprocessing, the next step is feature selection, where the key features associated with the research domain are prepared for classification. In feature selection, relevant subsets of features are selected so that they are able to predict mental disorder symptoms or accurately mark participants while bypassing overfitting. Analyzing statistics aims to identify parameters that differentiate non-disordered users from disordered users. Previously, Pearson correlation coefficients and Spearman rank correlation coefficients were applied (Park et al., 2015; Safa et al., 2022), as well as Mann-Whitney U tests (Ferwerda & Tkalcic, 2018; Park et al., 2013). PCA[28] (De Choudhury, Gamon, et al., 2013), gain ratio (Prieto et al., 2014), forward greedy stepwise (Hu et al., 2015), relief technique (Prieto et al., 2014), and convolutional neural network with cross-auto encoder technique (Lin, Jia, Guo, Xue, Li, et al., 2014) are also used to reduce the dimensionality of features. Finally, the model will be constructed using the training dataset, and the performance will be analogized and assessed. Consequently, the model can predict new user states (Unseen data) with gained accuracy. Figures 5A and 5B depict the overall survey and self-report-based analysis frameworks.

---

[27] https://docs.imagga.com
[28] Principal component analysis

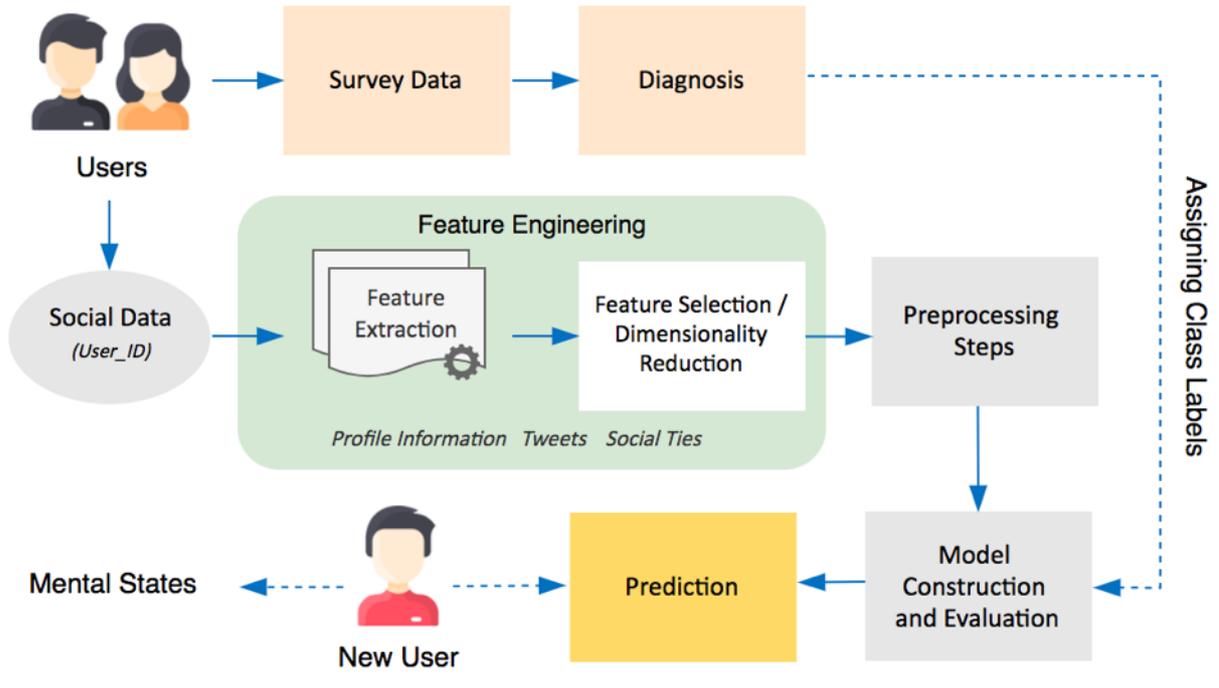

**A)** Survey-based analysis

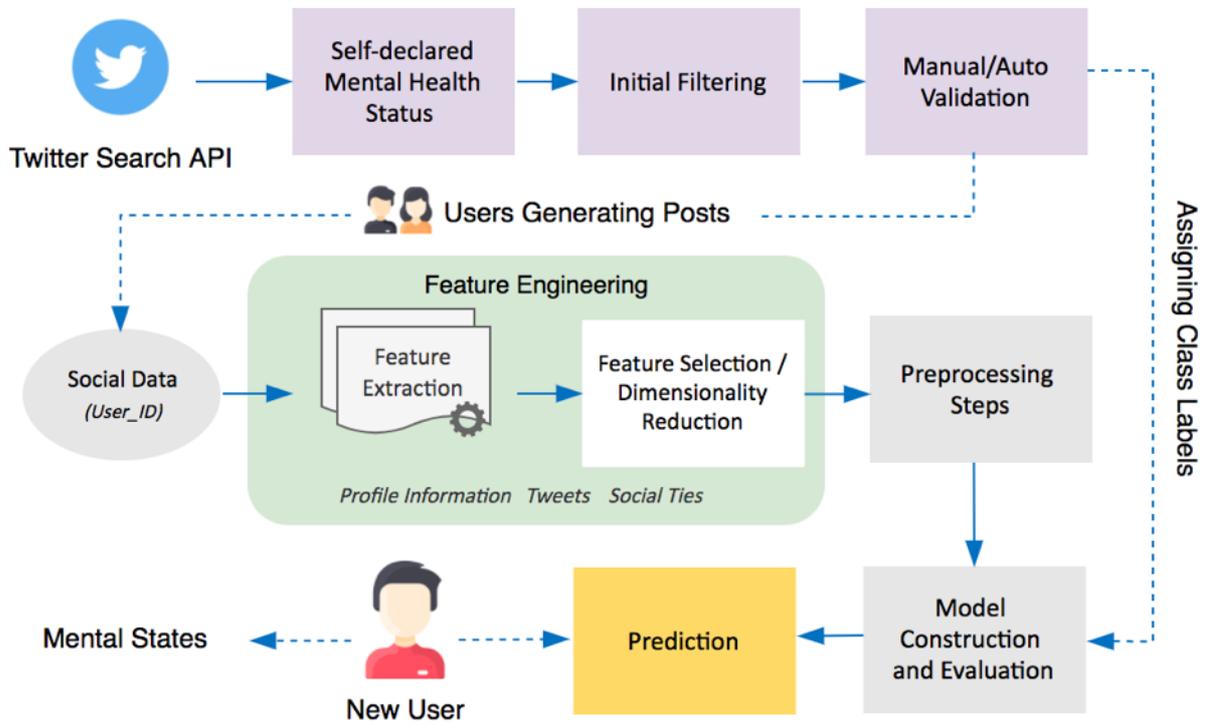

B) Self-report-based analysis

**Figure 5**. The overall framework of mental health prediction via social data

# 5. Prediction Algorithms

Machine learning algorithms learn patterns from data by using selected features as training sets. Five broad categories of machine learning techniques are presented here: supervised, semi-supervised, unsupervised, reinforcement, and deep learning.

## 5.1. Supervised learning

Supervised learning involves presenting algorithms with examples for training. Instances consist of inputs and outputs. Inputs are often represented as numerical arrays that describe examples numerically. The feature vector is usually referred to as this array. A classification problem results in an output called a class, a value we want to predict, such as a mental state. By training, the model can predict incognito output values from unexplored, unseen input values (Bonaccorso, 2017). The model can be tested for performance when the instances' output values are hidden. After that, the predictions can be compared to the ground truth, which is the actual output.

Classification refers to categorical output predictions; Regression refers to quantitative output predictions. A model can be trained using a set of signals from user behavior on social media with labels corresponding to mental states. Due to the categorical output (e.g., depressed and not depressed) a classifier can be applied. Therefore, most machine learning models used to detect mental states are classifiers. Allowing the categorical output variable to take a given desirable mental state as a value makes it possible to detect the presence/absence of a particular mental state. It is possible to determine the final category from probabilities rather than the final class from some classifiers. Cognitive state detection has been achieved through supervised learning classifiers such as Decision Tree (Burnap et al., 2015; Huang et al., 2014), Random Forest (Chen, Sykora, Jackson, Elayan, et al., 2018; Safa et al., 2022; Tadesse et al., 2019), AdaBoost (Tadesse et al., 2019), Support Vector Machines (SVM) Together with other kernels like Linear (Islam et al., 2018; Safa et al., 2022; Tadesse et al., 2019), and Radial Basis Function (RBF) (Burnap et al., 2015; De Choudhury, Counts, et al., 2013b; De Choudhury, Gamon, et al., 2013; Kang et al., 2016; Preoţiuc-Pietro et al., 2015; Tsugawa et al., 2015; Wang et al., 2017), Naïve Bayes (Burnap et al., 2015; Huang et al., 2014; Wang et al., 2017), different types of Regression (Coppersmith et al., 2014; Coppersmith et al., 2016; Hu et al., 2015; Nguyen et al., 2014; Yin et al., 2019), Artificial Neural Networks (Kim et al., 2020; Safa et al., 2022), etc. In Figure 6, a part of the survey conducted by Wongkoblap et al. (Wongkoblap et al., 2017) is visualized as a diagram to understand better how supervised learning methods classify users according to mental disorders are distributed in the literature.

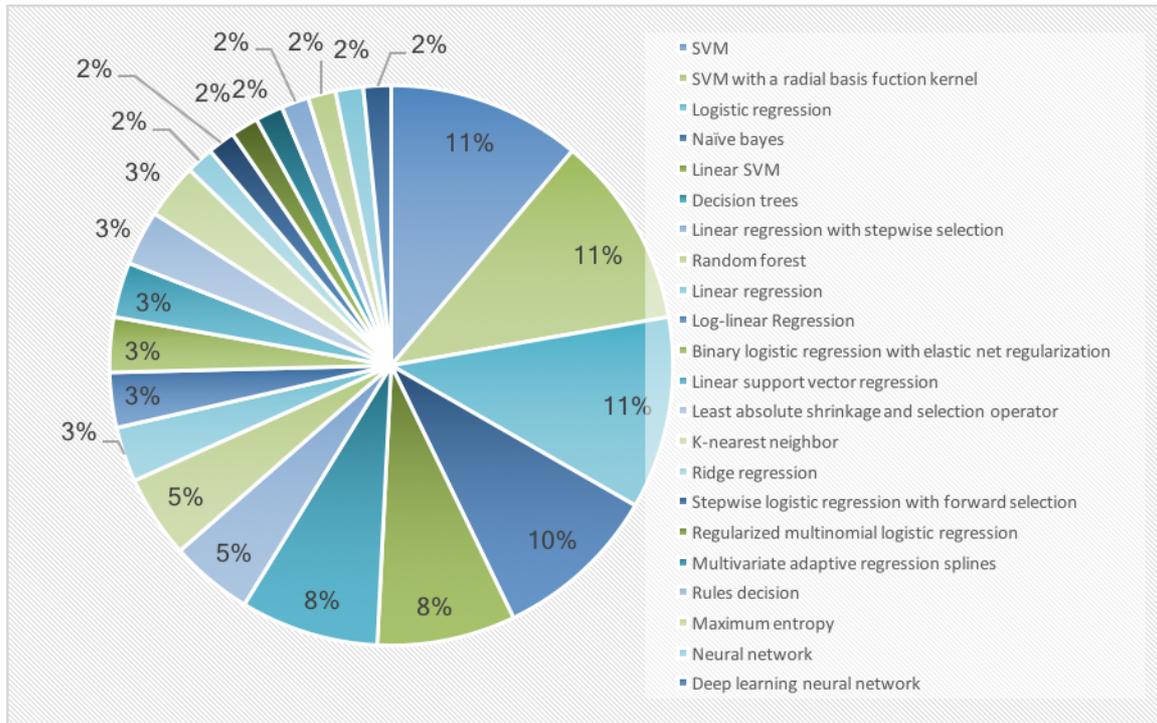

**Figure 6.** Distribution of supervised learning techniques in the selected articles based on Wongkoblap (Wongkoblap et al., 2017) research

## 5.2. Unsupervised learning

Unlike supervised learning, it is necessary to find the classes that can naturally arise from a similarity in input data instead of knowing the output variable at training time. Identifying groups of users with similar characteristics, for example. Unsupervised learning algorithms used for clustering are common methods for identifying groups or hierarchies within data. The unsupervised learning process can be considered a preprocessing step before the supervised method is applied. A number of standard clustering algorithms have been used for stress evaluation tasks (Garcia-Ceja et al., 2015; Xu et al., 2014), including k-medoids clustering, k-means clustering, hierarchical clustering, fuzzy c-means clustering, and density-based spatial clustering of applications with noise (DBSCAN) (Bonaccorso, 2017).

Finding a numerical textual content representation is crucial since clustering methods rely on numerical data as input. In more precise terms, the posts would have to be mapped into a latent space with an inherent structure based on contextual similarity. According to this continuous space, similar tweets have similar number vectors, and different tweets are distant. Embedding models can help solve this problem. Representing words as numerical vectors to capture the semantics can be equated to representing text as numerical vectors with meaning (Bayer et al., 2021). Using word embeddings, which are real-valued vectors, similar words are represented similarly. Embeddings of words have become an important trend in NLP research since the development of Word2Vec (Mikolov et al., 2013). A single sentence or document embedding can then be obtained by further processing the words of a sentence or document.

### 5.3. Semi-supervised learning

Semi-supervised learning occurs when a large number of training samples are available, but the output labels are only known for a small percentage of samples. Models are trained by applying labeled and unlabeled instances in semi-supervised algorithms (Zhu & Goldberg, 2009). Since ground truth classes are difficult to tag, semi-supervised learning helps detect mental states. For example, questionnaires are usually used to tag daily mood states; nevertheless, it is not uncommon for participants to forget to answer them, resulting in multifold days going untagged. Identifying the current state of a bipolar patient requires an expert's evaluation, usually clinical, to classify the data. This type of issue likely limits data tags.

According to Tariq et al. (Tariq et al., 2019), data extracted from Reddit can be used to classify those who suffer from chronic mental illnesses. The proposed method uses semi-supervised learning (Co-training) by incorporating widely used machine learning techniques. According to the experimental results, co-training-based classification appears more effective than typical approaches.

### 5.4. Reinforcement learning

Reinforcement learning occurs when agents learn through trial-and-error interactions with their environment. Depending on the current state, an agent will decide what action to take in order to maximize its rewards. A reinforcement learning algorithm explores a wide range of activities and chooses one that maximizes reward instead of being presented with inputs and targets (Bonaccorso, 2017). Agents are rewarded or punished when they perform the right or wrong actions. Over time, the system tries to figure out what actions result in the gilt-edged rewards. Its advantage is that no human expertise is required to understand the problem domain with reinforcement learning.

According to Gui et al. (Gui et al., 2019), depression can be detected using only contextual information by analyzing what users post. From users' recorded posts, they suggested a reinforcement learning-based approach for selecting indicator posts automatically. Compared to feature-based and neural network-based methods, the proposed method outperforms them.

### 5.5. Deep learning

Research on e-mental health has been boosted recently by deep learning, a rapidly-expanding field of machine learning. In contrast to statistical methods, deep learning uses neural networks, usually with many hidden layers, to learn various abstraction levels. The values of these layers are not included in the input data, making them referred to as hidden layers. It is, therefore, necessary for the network to determine which concepts help explain observed relationships. The MLP[29] architecture is used in some studies (Maupomé & Meurs, 2018). In contrast, more complex network architectures are used in others (Cho et al., 2014), including LSTM (Paul et al., 2018), GRU[30] (Sadeque et al., 2017), and CNN (Trotzek et al., 2018). Various RNNs[31], including LSTM and GRU, have been presented over the past few years, with the main difference being how the input maps to the output. They hold significant promise for detecting mental health conditions in clinical notes and social data because they demonstrate state-of-the-art implementation in plenty of applications, including NLP. It has been investigated whether deep learning could detect depression (Orabi et al., 2018; Wang et al., 2020) or recognize suicide-related psychiatric stressors (Du et al., 2018). A study by Gkotsis et al. (Gkotsis et al., 2017) analyzed posts on Reddit and developed classifiers for identifying and categorizing mental illness posts. In the dataset, they automatically recognized mental illness-related posts using neural networks and deep learning.

---
[29] Multilayer Perceptron
[30] Gated Recurrent Unit
[31] Recurrent neural networks

In RNN, input elements are prioritized using an attention mechanism, providing some interpretability of results (Su et al., 2020). An attention mechanism was included in a hierarchical RNN architecture introduced to predict the classes of posts, such as anxiety, depression, etc. (Ive et al., 2018). With the attention mechanism, the authors observed that the model effectively predicts risk text and extracts elements critical for decision-making. Users' mental health status can be inferred from the properties of the shared images, interactions, and relationships within the textual content. Researchers have increasingly endeavored to compound these two information types with textual content to develop predictive models in recent years. Using psychological and art theories, researchers analyzed text, image, and comment representations at low and middle levels (Lin, Jia, Guo, Xue, Huang, et al., 2014). Using CNN's hybrid model, they integrated post content and social interactions (Lin et al., 2017). Based on the results, It was found that stressed people tended to have less connected friends than those who weren't.

### 5.5.1. Transfer learning

Transfer learning refers to the process of using a model developed for a task to develop a model for another one. As an initial point for NLP and computer vision, using pre-trained algorithms is a popular approach in deep learning. As a result of the extended computing and time resources needed to build neural network models and the significant improvements they provide on related topics, this approach is prevalent in deep learning. Due to the lack of millions of labeled data points to train such complex models, data science is an advantageous field. Basically, By combining knowledge from similar or other areas, transfer learning allows us to learn novel things in a diverse domain without or with limited labeled sets (Pan & Yang, 2009). It is possible, for instance, to identify cats from images in the same way that dogs are recognized from images. A typical example is a system that understands how to differentiate whether an image contains signs of a particular disease based on the categories (Pogorelov et al., 2017). Besides overcoming the problem of insufficient labeled data, this method can also be used in conjunction with other methods. In particular, it seems to work quite well regarding image-related use cases. The results are not very good if the domains are too different. Moreover, most research on this topic has been limited to images and videos. It is important to note that transfer learning has the potential to address the challenge of reducing the amount of labeled data required by e-mental health applications.

# 6. Evaluation

Analyzing the generalizability of a trained model refers to its estimated performance on unseen data, which is the purpose of model evaluation. The train and test sets are two subs of a dataset that can be applied to estimate a model's generalization ability. Holdout validation uses a training set for model training, while a testing set is used to assess its performance. A random assignment is often made to a training or testing subset. Machine learning models often perform superbly if evaluated with the same data they were trained with, but they may not generalize well when evaluated with new data. As a result, holdout validation allows for better estimations of generalization since the trained model doesn't include information about sample sizes from the testing set.

Parameter tuning is required for some models; Continuously tuning such parameters based on the testing set performance leads to the model overfitting risk. This can be avoided by dividing the dataset into training, validation, and testing. Hence, models are built using training data, and parameters are tuned with validation data. Generalization performance is then assessed based on the testing set. Depending on the application, the three sets are typically split between 60% (train set), 20% (validation set), and 20% (test set) (Garcia-Ceja et al., 2018). When there is a large amount of data, holdout validation makes sense. Cross-validation with $k$ folds is preferred when data is limited. A random division of the data into $k$ equal-sized subsets is used in this method; After that, $k$ iterations are performed. One subset is used to test the model, while the remaining is used to train it. All iterations are averaged to determine the performance. The estimate's variance can be reduced as $k$ increases, but the computational load increases as well. Typically, $k$ is considered 10 in most cases. It is also known as LOOCV[32], where the number of samples equals $k$.

Another commonly used method in social media monitoring of mental health is to apply the training set, all other users' data, and the data from a new user as a testing set to assess the performance of a model. Users do not need to train data for this user-independent or general model (Garcia-Ceja & Brena, 2015). A classification model's reliability is assessed after prediction by an evaluation mechanism. The problem's type (regression, classification, etc.) determines the metrics to use. In the literature, the most applied metrics and visualization tools helping to examine the proposed models' performance are classification accuracy (proportion of correctly classified instances), precision (the positive predictive value), recall (the true positive rate or sensitivity), F1-score (the harmonic mean of the precision and recall), and ROC[33] curve (Safa et al., 2022). According to Table 1 (the confusion matrix), these metrics are as follows:

**Table 1.** Confusion Matrix

|  | Relevant | Non-Relevant |
|---|---|---|
| Retrieved | True Positive (TP) | False Positive (FP) |
| Not Retrieved | False Negative (FN) | True Negative (TN) |

---

[32] Leave-one-out cross-validation
[33] Receiver Operating Characteristics

1. $Accuracy = \frac{TP+TN}{TP+TN+FP+FN}$

2. $Precision = \frac{TP}{TP+FP}$

3. $Recall = TPR = \frac{TP}{TP+FN}$

4. $F1 = 2 \times \frac{Precision \times Recall}{Precision+Recall}$

ROC is a probability curve and AUC[34] measures separability. It determines the degree to which the model will be able to distinguish between different classes. Increasing the AUC increases the model's ability to predict 0 classes as 0 and 1 classes as 1. Analogously, users with the disorder and control group with a higher AUC can be distinguished more easily. ROC curves are plotted with TPR[35] versus FPR[36], which are computed as follows. Specificity refers to the true negative rate, i.e., how many negatives are categorized precisely.

5. $Specificity = \frac{TN}{TN+FP}$

6. $FPR = 1 - Specificity = \frac{FP}{TN+FP}$

In evaluating performance, it is essential to avoid using only one metric. When several different metrics are used simultaneously, a classifier or prediction approach's real performance and robustness can be better assessed (Garcia-Ceja & Brena, 2015). As a result, reporting all metrics is good practice when classes are imbalanced, as precision may be deceptive. A sample imbalance occurs when only a tiny percentage of samples belong to a particular category.

Typical performance metrics for regression problems include mean square, root means square, and mean absolute error, as well as correlation coefficients (Botchkarev, 2018). Mental state detection often produces ordinal output classes. In other words, it has a natural order; It is possible to model depression levels with an ordinal variable that takes the values " high," "medium," and "low." In regular performance metrics, errors are treated equally: Confusion between low and medium has equivalent error weight as confusion between low and high; obviously, the latter mistake should be penalized intensively. Metrics like linear correlation, mean squared or absolute errors, and accuracy within $n$ can be used to measure ordinal variables.

---

[34] Area Under the ROC Curve
[35] True Positive Rate
[36] False Positive Rate

## 7. Status, challenges, and future direction

In mental health diagnostics and classifications, mental health professionals use DSM[37] and the ICD[38] to classify mental disorders based on their type, intensity, and duration (Statistics & Administration, 1980). Although these tools are frequently used, they have certain limitations. Many diseases are diagnosed using the same criteria; it is, therefore, impossible to distinguish between diagnostic groups. Further, these tools do not take into account additional factors like demographics and biochemistry, patient interview records, mental disorder background in the family, and medication response (Tutun et al., 2022). In spite of the fact that screening surveys that have been validated and are reliable are closest to clinical practice, they are time-consuming to administer and rely on the self-selection of crowd workers, introducing biases in the sampling (Arditte et al., 2016). The assessments of this approach also suffer from significant temporal gaps, as identifying risk factors for mental illness often requires immediate action, limiting the development of effective interventions (De Choudhury, 2013). It is also possible that mentally ill people are less likely to cooperate with researchers; Social media analysis can detect cases that would otherwise go undiagnosed. In contrast, public data approaches (i.e. selecting users based on self-report analysis, or specific forum membership) have a larger sample size and bring additional information. As of yet, no studies have focused explicitly on effectively recognizing someone oblivious to their mental health condition, so the chances of capturing users unaware of their diagnosis are low (Guntuku et al., 2017).

For social media-based screening to be effective and distinguish between mental illnesses, gold-standard structured clinical interviews, and further screening methods must be combined with social media data collection in ecologically valid samples (Coppersmith, Dredze, Harman, Hollingshead et al., 2015). It remains challenging to gather, process, fusion, and analyze big social data from unstructured, semi-structured sources to exploit valuable knowledge. Due to this issue, several challenges and problems have emerged in the social big data domain, including knowledge representation, data processing, data analysis, data management, and data visualization (Bello-Orgaz et al., 2016; Kumar et al., 2020).

However, most former works investigated textual/visual features from social media platforms without considering users' social networks. Mental health problems are highly associated with friends' circles, suggesting that social network analysis may be an effective method for studying the prevalence of such disorders (Rosenquist et al., 2011). It remains challenging, however, to model text information and network structure comprehensively. A convolutional graph network has been developed to address networked data mining in this context.

In practice, the final prediction model is often derived by combining different types of algorithms. Before building supervised learning models, for insistence, unsupervised learning models are often applied as a preprocessing phase (Garcia-Ceja et al., 2018). It is also important to differentiate between user-dependent and user-independent (known as general) training schemes. Training the former is founded on the data from the particular user under consideration. All other users' data are used to train the latter, except the target user (who is using the system). User-dependent models are advantageous since they capture the behavior of each individual user and produce better results, but they require a substantial amount of training time. Users who are 'atypical' might not benefit from user-independent models since they need no data from the target user.

---

[37] The Diagnostic and Statistical Manual of Mental Disorders
[38] International Classification of Diseases

Although social media analyses are applied to detect online users with mental disorders, the importance of translating this innovation into practical application and providing users with real-time assistance, for instance, cannot be overstated (Rice et al., 2014). In addition, the lack of large training data is one of the main obstacles to the widespread application of deep learning approaches in e-mental health. It is worth noting that the training data used for these deep architectures is extensive and hand-labeled.

There are some ethical concerns associated with social-media-based mental illness assessment. Privacy has been a concern for a long time. People with mental illness may be disadvantaged by these policies when their employers or insurers use them. A data protection and ownership framework are essential for ensuring that mental illnesses are not stigmatized or discriminated against (Luna Ansari et al.). Most people are unaware that their digital traces contain mental-health-related information. Transparency is essential regarding who generates which health indicators and why. Mandatory reporting guidelines need to be clarified from a mental health perspective. Because of misclassifications, derived mental health indicators are difficult to integrate into systems of care responsibly (Guntuku et al., 2017).

## 8. Conclusion

Social networks are becoming increasingly popular as a platform for sharing opinions, feelings, and thoughts. Recently, social networks have been exploited as a public health tool with increasing interest. This chapter discussed the implications related to big social data and explained how these data could be used to analyze and predict mental disorders. We reviewed the recent approaches addressing mental state evaluation and disorders diagnosis using users' digital footprints. The studies, as mentioned earlier, were organized concerning the assessment strategy, data configuration (including data preprocessing and the feature engineering phase), and prediction algorithms. Furthermore, we presented a series that examines the language and behavior of people enduring mental disorders (like PTSD or depression) and discussed diverse aspects associated with developing experimental frameworks. Thus, this chapter's primary contributions are comprehensively analyzing mental state assessment methods on social data, structurally categorizing them by considering their design principles and lessons learned during their development, and discussing challenges and achievable directions for future studies.

**References**

Arditte, K. A., Çek, D., Shaw, A. M., & Timpano, K. R. (2016). The importance of assessing clinical phenomena in Mechanical Turk research. *Psychological assessment, 28*(6), 684.

Association, A. P. (2013). *Diagnostic and statistical manual of mental disorders (DSM-5®)*. American Psychiatric Pub.

Balcombe, L., & De Leo, D. (2021). Digital mental health challenges and the horizon ahead for solutions. *JMIR mental health, 8*(3), e26811.

Bayer, M., Kaufhold, M.-A., & Reuter, C. (2021). Information Overload in Crisis Management: Bilingual Evaluation of Embedding Models for Clustering Social Media Posts in Emergencies. ECIS.

Bello-Orgaz, G., Jung, J. J., & Camacho, D. (2016). Social big data: Recent achievements and new challenges. *Information Fusion, 28*, 45-59.

Bersani, F. S., Barchielli, B., Ferracuti, S., Panno, A., Carbone, G. A., Massullo, C., Farina, B., Corazza, O., Prevete, E., & Tarsitani, L. (2022). The association of problematic use of social media and online videogames with aggression is mediated by insomnia severity: A cross-sectional study in a sample of 18-to 24-year-old individuals. *Aggressive behavior, 48*(3), 348-355.

Blei, D. M., Ng, A. Y., & Jordan, M. I. (2003). Latent dirichlet allocation. *Journal of machine Learning research, 3*(Jan), 993-1022.

Boettcher, N. (2021). Studies of Depression and Anxiety Using Reddit as a Data Source: Scoping Review. *JMIR mental health, 8*(11), e29487.

Bollen, J., Gonçalves, B., Ruan, G., & Mao, H. (2011). Happiness is assortative in online social networks. *Artificial life, 17*(3), 237-251.

Bonaccorso, G. (2017). *Machine learning algorithms*. Packt Publishing Ltd.

Botchkarev, A. (2018). Performance metrics (error measures) in machine learning regression, forecasting and prognostics: Properties and typology. *arXiv preprint arXiv:1809.03006*.

Braithwaite, S. R., Giraud-Carrier, C., West, J., Barnes, M. D., & Hanson, C. L. (2016a). Validating machine learning algorithms for Twitter data against established measures of suicidality. *JMIR mental health, 3*(2), e4822.

Braithwaite, S. R., Giraud-Carrier, C., West, J., Barnes, M. D., & Hanson, C. L. (2016b). Validating machine learning algorithms for Twitter data against established measures of suicidality. *JMIR mental health, 3*(2), e21.

Burnap, P., Colombo, W., & Scourfield, J. (2015). Machine classification and analysis of suicide-related communication on twitter. Proceedings of the 26th ACM conference on hypertext & social media.

Cavazos-Rehg, P. A., Krauss, M. J., Sowles, S., Connolly, S., Rosas, C., Bharadwaj, M., & Bierut, L. J. (2016). A content analysis of depression-related tweets. *Computers in human behavior, 54*, 351-357.

Chen, X., Sykora, M., Jackson, T., Elayan, S., & Munir, F. (2018). Tweeting Your Mental Health: an Exploration of Different Classifiers and Features with Emotional Signals in Identifying Mental Health Conditions.

Chen, X., Sykora, M. D., Jackson, T. W., & Elayan, S. (2018). What about mood swings: Identifying depression on twitter with temporal measures of emotions. Companion Proceedings of the The Web Conference 2018.

Chiu, C. Y., Lane, H. Y., Koh, J. L., & Chen, A. L. (2020). Multimodal depression detection on instagram considering time interval of posts. *Journal of Intelligent Information Systems*, 1-23.

Cho, K., Van Merriënboer, B., Gulcehre, C., Bahdanau, D., Bougares, F., Schwenk, H., & Bengio, Y. (2014). Learning phrase representations using RNN encoder-decoder for statistical machine translation. *arXiv preprint arXiv:1406.1078*.

Chua, C. E. H., Storey, V. C., Li, X., & Kaul, M. (2019). Developing insights from social media using semantic lexical chains to mine short text structures. *Decision Support Systems, 127*, 113142.

Conway, M., & O'Connor, D. (2016). Social media, big data, and mental health: current advances and ethical implications. *Current opinion in psychology, 9*, 77-82.

Coppersmith, G., Dredze, M., & Harman, C. (2014). Quantifying mental health signals in Twitter. Proceedings of the workshop on computational linguistics and clinical psychology: From linguistic signal to clinical reality.

Coppersmith, G., Dredze, M., Harman, C., & Hollingshead, K. (2015). From ADHD to SAD: Analyzing the language of mental health on Twitter through self-reported diagnoses. Proceedings of the 2nd workshop on computational linguistics and clinical psychology: from linguistic signal to clinical reality.

Coppersmith, G., Dredze, M., Harman, C., Hollingshead, K., & Mitchell, M. (2015). CLPsych 2015 shared task: Depression and PTSD on Twitter. Proceedings of the 2nd workshop on computational linguistics and clinical psychology: from linguistic signal to clinical reality.

Coppersmith, G., Leary, R., Crutchley, P., & Fine, A. (2018). Natural language processing of social media as screening for suicide risk. *Biomedical informatics insights, 10*, 1178222618792860.

Coppersmith, G., Ngo, K., Leary, R., & Wood, A. (2016). Exploratory analysis of social media prior to a suicide attempt. Proceedings of the Third Workshop on Computational Linguistics and Clinical Psychology.

Cortes, C., & Vapnik, V. (1995). Support-vector networks. *Machine learning, 20*(3), 273-297.

Crone, E. A., & Konijn, E. A. (2018). Media use and brain development during adolescence. *Nature communications, 9*(1), 1-10.

De Choudhury, M. (2013). Role of social media in tackling challenges in mental health. Proceedings of the 2nd international workshop on Socially-aware multimedia.

De Choudhury, M., Counts, S., & Horvitz, E. (2013a). Major life changes and behavioral markers in social media: case of childbirth. Proceedings of the 2013 conference on Computer supported cooperative work.

De Choudhury, M., Counts, S., & Horvitz, E. (2013b). Social media as a measurement tool of depression in populations. Proceedings of the 5th Annual ACM Web Science Conference.

De Choudhury, M., Counts, S., Horvitz, E. J., & Hoff, A. (2014). Characterizing and predicting postpartum depression from shared facebook data. Proceedings of the 17th ACM conference on Computer supported cooperative work & social computing.

De Choudhury, M., Gamon, M., Counts, S., & Horvitz, E. (2013). Predicting depression via social media. *Icwsm, 13*, 1-10.

De Choudhury, M., Kiciman, E., Dredze, M., Coppersmith, G., & Kumar, M. (2016). Discovering shifts to suicidal ideation from mental health content in social media. Proceedings of the 2016 CHI conference on human factors in computing systems.

Docrat, S., Besada, D., Cleary, S., Daviaud, E., & Lund, C. (2019). Mental health system costs, resources and constraints in South Africa: a national survey. *Health policy and planning, 34*(9), 706-719.

Du, J., Zhang, Y., Luo, J., Jia, Y., Wei, Q., Tao, C., & Xu, H. (2018). Extracting psychiatric stressors for suicide from social media using deep learning. *BMC medical informatics and decision making, 18*(2), 43.

Durahim, A. O., & Coşkun, M. (2015). # iamhappybecause: Gross National Happiness through Twitter analysis and big data. *Technological Forecasting and Social Change, 99*, 92-105.

Ebert, D. D., Harrer, M., Apolinário-Hagen, J., & Baumeister, H. (2019). Digital interventions for mental disorders: key features, efficacy, and potential for artificial intelligence applications. In *Frontiers in Psychiatry* (pp. 583-627). Springer.

Ferwerda, B., & Tkalcic, M. (2018). You are what you post: What the content of Instagram pictures tells about users' personality. The 23rd International on Intelligent User Interfaces, March 7-11, Tokyo, Japan.

Garcia-Ceja, E., & Brena, R. (2015). Building personalized activity recognition models with scarce labeled data based on class similarities. International conference on ubiquitous computing and ambient intelligence.

Garcia-Ceja, E., Osmani, V., & Mayora, O. (2015). Automatic stress detection in working environments from smartphones' accelerometer data: a first step. *IEEE journal of biomedical and health informatics, 20*(4), 1053-1060.

Garcia-Ceja, E., Riegler, M., Nordgreen, T., Jakobsen, P., Oedegaard, K. J., & Tørresen, J. (2018). Mental health monitoring with multimodal sensing and machine learning: A survey. *Pervasive and Mobile Computing, 51*, 1-26.

Garg, M., Saxena, C., Krishnan, V., Joshi, R., Saha, S., Mago, V., & Dorr, B. J. (2022). CAMS: An Annotated Corpus for Causal Analysis of Mental Health Issues in Social Media Posts. *arXiv preprint arXiv:2207.04674*.

Gkotsis, G., Oellrich, A., Velupillai, S., Liakata, M., Hubbard, T. J., Dobson, R. J., & Dutta, R. (2017). Characterisation of mental health conditions in social media using Informed Deep Learning. *Scientific reports, 7*(1), 1-11.

Gui, T., Zhang, Q., Zhu, L., Zhou, X., Peng, M., & Huang, X. (2019). Depression detection on social media with reinforcement learning. China National Conference on Chinese Computational Linguistics.

Guntuku, S. C., Preotiuc-Pietro, D., Eichstaedt, J. C., & Ungar, L. H. (2019). What twitter profile and posted images reveal about depression and anxiety. Proceedings of the international AAAI conference on web and social media.

Guntuku, S. C., Yaden, D. B., Kern, M. L., Ungar, L. H., & Eichstaedt, J. C. (2017). Detecting depression and mental illness on social media: an integrative review. *Current Opinion in Behavioral Sciences, 18*, 43-49.

Hanna, F., Barbui, C., Dua, T., Lora, A., van Regteren Altena, M., & Saxena, S. (2018). Global mental health: how are we doing? *World Psychiatry, 17*(3), 367.

Haque, A., Reddi, V., & Giallanza, T. (2021). Deep learning for suicide and depression identification with unsupervised label correction. International Conference on Artificial Neural Networks.

Harvey, D., Lobban, F., Rayson, P., Warner, A., & Jones, S. (2022). Natural Language Processing Methods and Bipolar Disorder: Scoping Review. *JMIR mental health, 9*(4), e35928.

Hu, Q., Li, A., Heng, F., Li, J., & Zhu, T. (2015). Predicting depression of social media user on different observation windows. 2015 IEEE/WIC/ACM International Conference on Web Intelligence and Intelligent Agent Technology (WI-IAT).

Huang, X., Zhang, L., Chiu, D., Liu, T., Li, X., & Zhu, T. (2014). Detecting suicidal ideation in Chinese microblogs with psychological lexicons. 2014 IEEE 11th Intl Conf on Ubiquitous Intelligence and Computing and 2014 IEEE 11th Intl Conf on Autonomic and Trusted Computing and 2014 IEEE 14th Intl Conf on Scalable Computing and Communications and Its Associated Workshops.

Islam, M. R., Kabir, M. A., Ahmed, A., Kamal, A. R. M., Wang, H., & Ulhaq, A. (2018). Depression detection from social network data using machine learning techniques. *Health information science and systems, 6*(1), 1-12.

Ive, J., Gkotsis, G., Dutta, R., Stewart, R., & Velupillai, S. (2018). Hierarchical neural model with attention mechanisms for the classification of social media text related to mental health. Proceedings of the Fifth Workshop on Computational Linguistics and Clinical Psychology: From Keyboard to Clinic.

Javadi, S., Safa, R., Azizi, M., & Mirroshandel, S. A. (2020). A Recommendation System for Finding Experts in Online Scientific Communities. *Journal of AI and Data Mining, 8*(4), 573-584.

Ji, S., Yu, C. P., Fung, S.-f., Pan, S., & Long, G. (2018). Supervised learning for suicidal ideation detection in online user content. *Complexity, 2018*.

Johnson, M., Albizri, A., Harfouche, A., & Tutun, S. (2021). Digital transformation to mitigate emergency situations: increasing opioid overdose survival rates through explainable artificial intelligence. *Industrial Management & Data Systems*.

Kang, K., Yoon, C., & Kim, E. Y. (2016). Identifying depressive users in Twitter using multimodal analysis. 2016 International Conference on Big Data and Smart Computing (BigComp).

Kern, M. L., Park, G., Eichstaedt, J. C., Schwartz, H. A., Sap, M., Smith, L. K., & Ungar, L. H. (2016). Gaining insights from social media language: Methodologies and challenges. *Psychological methods, 21*(4), 507.

Kilbourne, A. M., Beck, K., Spaeth-Rublee, B., Ramanuj, P., O'Brien, R. W., Tomoyasu, N., & Pincus, H. A. (2018). Measuring and improving the quality of mental health care: a global perspective. *World Psychiatry, 17*(1), 30-38.

Kim, J., Lee, J., Park, E., & Han, J. (2020). A deep learning model for detecting mental illness from user content on social media. *Scientific reports, 10*(1), 1-6.

Kiritchenko, S., Zhu, X., & Mohammad, S. M. (2014). Sentiment analysis of short informal texts. *Journal of Artificial Intelligence Research, 50*, 723-762.

Krieck, M., Dreesman, J., Otrusina, L., & Denecke, K. (2011). A new age of public health: Identifying disease outbreaks by analyzing tweets. Proceedings of health web-science workshop, ACM Web Science Conference.

Kumar, A., & Garg, G. (2019). Sentiment analysis of multimodal twitter data. *Multimedia Tools and Applications, 78*(17), 24103-24119.

Kumar, A., Sangwan, S. R., & Nayyar, A. (2020). Multimedia social big data: Mining. In *Multimedia big data computing for IoT applications* (pp. 289-321). Springer.

Lee, P., Abernethy, A., Shaywitz, D., Gundlapalli, A. V., Weinstein, J., Doraiswamy, P. M., Schulman, K., & Madhavan, S. (2022). Digital Health COVID-19 Impact Assessment: Lessons Learned and Compelling Needs. *NAM perspectives, 2022*.

Lin, H., Jia, J., Guo, Q., Xue, Y., Huang, J., Cai, L., & Feng, L. (2014). Psychological stress detection from cross-media microblog data using deep sparse neural network. 2014 IEEE International Conference on Multimedia and Expo (ICME).

Lin, H., Jia, J., Guo, Q., Xue, Y., Li, Q., Huang, J., Cai, L., & Feng, L. (2014). User-level psychological stress detection from social media using deep neural network. Proceedings of the 22nd ACM international conference on Multimedia.

Lin, H., Jia, J., Qiu, J., Zhang, Y., Shen, G., Xie, L., Tang, J., Feng, L., & Chua, T.-S. (2017). Detecting stress based on social interactions in social networks. *IEEE Transactions on Knowledge and Data Engineering, 29*(9), 1820-1833.

Lopez-Castroman, J., Moulahi, B., Azé, J., Bringay, S., Deninotti, J., Guillaume, S., & Baca-Garcia, E. (2020). Mining social networks to improve suicide prevention: A scoping review. *Journal of neuroscience research, 98*(4), 616-625.

Losada, D. E., & Crestani, F. (2016). A test collection for research on depression and language use. International Conference of the Cross-Language Evaluation Forum for European Languages.

Losada, D. E., Crestani, F., & Parapar, J. (2020). eRisk 2020: Self-harm and Depression Challenges. European Conference on Information Retrieval.

Loveys, K., Crutchley, P., Wyatt, E., & Coppersmith, G. (2017). Small but mighty: affective micropatterns for quantifying mental health from social media language. Proceedings of the Fourth Workshop on Computational Linguistics and Clinical Psychology—From Linguistic Signal to Clinical Reality.

Luna Ansari, S. J., Chen, Q., Cambria, E., & Pekka, P. M. Ensemble Hybrid Learning Methods for Automated Depression Detection.

Lv, M., Li, A., Liu, T., & Zhu, T. (2015). Creating a Chinese suicide dictionary for identifying suicide risk on social media. *PeerJ, 3*, e1455.

Ma, L., Wang, Z., & Zhang, Y. (2017). Extracting depression symptoms from social networks and web blogs via text mining. International Symposium on Bioinformatics Research and Applications.

Mao, K., Niu, J., Chen, H., Wang, L., & Atiquzzaman, M. (2018). Mining of marital distress from microblogging social networks: A case study on Sina Weibo. *Future Generation Computer Systems, 86*, 1481-1490.

Margus, C., Brown, N., Hertelendy, A. J., Safferman, M. R., Hart, A., & Ciottone, G. R. (2021). Emergency physician Twitter use in the COVID-19 pandemic as a potential predictor of impending surge: Retrospective observational study. *Journal of medical Internet research, 23*(7), e28615.

Martínez-Castaño, R., Pichel, J. C., & Losada, D. E. (2020). A Big Data Platform for Real Time Analysis of Signs of Depression in Social Media. *International Journal of Environmental Research and Public Health, 17*(13), 4752.

Maupomé, D., & Meurs, M.-J. (2018). Using Topic Extraction on Social Media Content for the Early Detection of Depression. *CLEF (Working Notes), 2125*.

Mikolov, T., Chen, K., Corrado, G., & Dean, J. (2013). Efficient estimation of word representations in vector space. *arXiv preprint arXiv:1301.3781*.

Mowery, D. L., Bryan, C., & Conway, M. (2015). Towards developing an annotation scheme for depressive disorder symptoms: A preliminary study using twitter data. Proceedings of the 2nd Workshop on Computational Linguistics and Clinical Psychology: From Linguistic Signal to Clinical Reality.

Neter, J., Kutner, M. H., Nachtsheim, C. J., & Wasserman, W. (1996). Applied linear statistical models.

Nguyen, T., Phung, D., Dao, B., Venkatesh, S., & Berk, M. (2014). Affective and content analysis of online depression communities. *IEEE Transactions on Affective Computing, 5*(3), 217-226.

Orabi, A. H., Buddhitha, P., Orabi, M. H., & Inkpen, D. (2018). Deep learning for depression detection of twitter users. Proceedings of the Fifth Workshop on Computational Linguistics and Clinical Psychology: From Keyboard to Clinic.

Organization, W. H. (2021). Suicide worldwide in 2019: global health estimates.

Pan, S. J., & Yang, Q. (2009). A survey on transfer learning. *IEEE Transactions on Knowledge and Data Engineering, 22*(10), 1345-1359.

Park, S., Kim, I., Lee, S. W., Yoo, J., Jeong, B., & Cha, M. (2015). Manifestation of depression and loneliness on social networks: a case study of young adults on Facebook. Proceedings of the 18th ACM conference on computer supported cooperative work & social computing.

Park, S., Lee, S. W., Kwak, J., Cha, M., & Jeong, B. (2013). Activities on Facebook reveal the depressive state of users. *Journal of medical Internet research, 15*(10), e2718.

Patel, V., Saxena, S., Lund, C., Thornicroft, G., Baingana, F., Bolton, P., Chisholm, D., Collins, P. Y., Cooper, J. L., & Eaton, J. (2018). The Lancet Commission on global mental health and sustainable development. *The lancet, 392*(10157), 1553-1598.

Paul, M. J., & Dredze, M. (2011). You are what you tweet: Analyzing twitter for public health. Fifth international AAAI conference on weblogs and social media.

Paul, S., Jandhyala, S. K., & Basu, T. (2018). Early Detection of Signs of Anorexia and Depression Over Social Media using Effective Machine Learning Frameworks. CLEF (Working Notes).

Pedersen, T. (2015). Screening Twitter users for depression and PTSD with lexical decision lists. Proceedings of the 2nd workshop on computational linguistics and clinical psychology: from linguistic signal to clinical reality.

Pennebaker, J. W., Boyd, R. L., Jordan, K., & Blackburn, K. (2015). *The development and psychometric properties of LIWC2015*.

Petersen, I., Bhana, A., Fairall, L. R., Selohilwe, O., Kathree, T., Baron, E. C., Rathod, S. D., & Lund, C. (2019). Evaluation of a collaborative care model for integrated primary care of common mental disorders comorbid with chronic conditions in South Africa. *BMC psychiatry, 19*(1), 1-11.

Plaza-del-Arco, F. M., Martín-Valdivia, M. T., Ureña-López, L. A., & Mitkov, R. (2020). Improved emotion recognition in Spanish social media through incorporation of lexical knowledge. *Future Generation Computer Systems, 110*, 1000-1008.

Pogorelov, K., Riegler, M., Eskeland, S. L., de Lange, T., Johansen, D., Griwodz, C., Schmidt, P. T., & Halvorsen, P. (2017). Efficient disease detection in gastrointestinal videos–global features versus neural networks. *Multimedia Tools and Applications, 76*(21), 22493-22525.

Preoţiuc-Pietro, D., Sap, M., Schwartz, H. A., & Ungar, L. (2015). Mental illness detection at the World Well-Being Project for the CLPsych 2015 shared task. Proceedings of the 2nd Workshop on Computational Linguistics and Clinical Psychology: From Linguistic Signal to Clinical Reality.

Prieto, V. M., Matos, S., Alvarez, M., Cacheda, F., & Oliveira, J. L. (2014). Twitter: a good place to detect health conditions. *PloS one, 9*(1), e86191.

Reece, A. G., & Danforth, C. M. (2017). Instagram photos reveal predictive markers of depression. *EPJ Data Science, 6*(1), 1-12.

Resnik, P., Armstrong, W., Claudino, L., Nguyen, T., Nguyen, V.-A., & Boyd-Graber, J. (2015). Beyond LDA: exploring supervised topic modeling for depression-related language in Twitter. Proceedings of the 2nd workshop on computational linguistics and clinical psychology: from linguistic signal to clinical reality.

Rice, S., Robinson, J., Bendall, S., Hetrick, S., Cox, G., Bailey, E., Gleeson, J., & Alvarez-Jimenez, M. (2016). Online and social media suicide prevention interventions for young people: a focus on implementation and moderation. *Journal of the Canadian Academy of Child and Adolescent Psychiatry, 25*(2), 80.

Rice, S. M., Goodall, J., Hetrick, S. E., Parker, A. G., Gilbertson, T., Amminger, G. P., Davey, C. G., McGorry, P. D., Gleeson, J., & Alvarez-Jimenez, M. (2014). Online and social networking interventions for the treatment of depression in young people: a systematic review. *Journal of medical Internet research, 16*(9), e3304.

Ríssola, E. A., Aliannejadi, M., & Crestani, F. (2020). Beyond Modelling: Understanding Mental Disorders in Online Social Media. European Conference on Information Retrieval,

Ríssola, E. A., Losada, D. E., & Crestani, F. (2021). A survey of computational methods for online mental state assessment on social media. *ACM Transactions on Computing for Healthcare, 2*(2), 1-31.

Rosenquist, J. N., Fowler, J. H., & Christakis, N. A. (2011). Social network determinants of depression. *Molecular psychiatry, 16*(3), 273-281.

Sadeque, F., Xu, D., & Bethard, S. (2017). UArizona at the CLEF eRisk 2017 pilot task: linear and recurrent models for early depression detection. CEUR workshop proceedings.

Safa, R., Bayat, P., & Moghtader, L. (2022). Automatic detection of depression symptoms in twitter using multimodal analysis. *The Journal of Supercomputing, 78*(4), 4709-4744.

Schwartz, H. A., Sap, M., Kern, M. L., Eichstaedt, J. C., Kapelner, A., Agrawal, M., Blanco, E., Dziurzynski, L., Park, G., & Stillwell, D. (2016). Predicting individual well-being through the language of social media. Biocomputing 2016: Proceedings of the Pacific Symposium.

Simonyan, K., & Zisserman, A. (2015). Very deep convolutional networks for large-scale image recognition. International Conference on Learning Representations.

Sooknanan, J., & Mays, N. (2021). Harnessing social media in the modelling of pandemics—challenges and opportunities. *Bulletin of Mathematical Biology, 83*(5), 1-11.

Spiro, E. S. (2016). Research opportunities at the intersection of social media and survey data. *Current Opinion in Psychology, 9*, 67-71.

Statistics, N. C. f. H., & Administration, U. S. H. C. F. (1980). *The International Classification of Diseases, 9th Revision, Clinical Modification: Diseases: alphabetic index* (Vol. 2). US Department of Health and Human Services, Public Health Service.

Su, C., Xu, Z., Pathak, J., & Wang, F. (2020). Deep learning in mental health outcome research: a scoping review. *Translational Psychiatry, 10*(1), 1-26.

Tadesse, M. M., Lin, H., Xu, B., & Yang, L. (2019). Detection of depression-related posts in reddit social media forum. *IEEE Access, 7*, 44883-44893.

Tariq, S., Akhtar, N., Afzal, H., Khalid, S., Mufti, M. R., Hussain, S., Habib, A., & Ahmad, G. (2019). A novel co-training-based approach for the classification of mental illnesses using social media posts. *IEEE Access, 7*, 166165-166172.

Thelwall, M., Buckley, K., Paltoglou, G., Cai, D., & Kappas, A. (2010). Sentiment strength detection in short informal text. *Journal of the American society for information science and technology, 61*(12), 2544-2558.

Thieme, A., Belgrave, D., & Doherty, G. (2020). Machine learning in mental health: A systematic review of the HCI literature to support the development of effective and implementable ML systems. *ACM Transactions on Computer-Human Interaction (TOCHI), 27*(5), 1-53.

Thorstad, R., & Wolff, P. (2019). Predicting future mental illness from social media: A big-data approach. *Behavior research methods, 51*(4), 1586-1600.

Trotzek, M., Koitka, S., & Friedrich, C. M. (2018). Word embeddings and linguistic metadata at the CLEF 2018 tasks for early detection of depression and anorexia. CLEF (Working Notes).

Tsugawa, S., Kikuchi, Y., Kishino, F., Nakajima, K., Itoh, Y., & Ohsaki, H. (2015). Recognizing depression from twitter activity. Proceedings of the 33rd annual ACM conference on human factors in computing systems.

Tutun, S., Johnson, M. E., Ahmed, A., Albizri, A., Irgil, S., Yesilkaya, I., Ucar, E. N., Sengun, T., & Harfouche, A. (2022). An AI-based Decision Support System for Predicting Mental Health Disorders. *Information Systems Frontiers*, 1-16.

Uban, A.-S., Chulvi, B., & Rosso, P. (2021). An emotion and cognitive based analysis of mental health disorders from social media data. *Future Generation Computer Systems, 124*, 480-494.

Wang, T., Brede, M., Ianni, A., & Mentzakis, E. (2017). Detecting and characterizing eating-disorder communities on social media. Proceedings of the Tenth ACM International conference on web search and data mining.

Wang, Y., Wang, Z., Li, C., Zhang, Y., & Wang, H. (2020). A Multitask Deep Learning Approach for User Depression Detection on Sina Weibo. *arXiv preprint arXiv:2008.11708*.

Whiteford, H., Ferrari, A., & Degenhardt, L. (2016). Global burden of disease studies: implications for mental and substance use disorders. *Health Affairs, 35*(6), 1114-1120.

Wilson, T., Hoffmann, P., Somasundaran, S., Kessler, J., Wiebe, J., Choi, Y., Cardie, C., Riloff, E., & Patwardhan, S. (2005). OpinionFinder: A system for subjectivity analysis. Proceedings of HLT/EMNLP 2005 Interactive Demonstrations.

Wongkoblap, A. (2020). Multiple Instance Learning for Detecting Depression Markers in Social Media Content.

Wongkoblap, A., Vadillo, M. A., & Curcin, V. (2017). Researching mental health disorders in the era of social media: systematic review. *Journal of medical Internet research, 19*(6), e7215.

Xu, Q., Nwe, T. L., & Guan, C. (2014). Cluster-based analysis for personalized stress evaluation using physiological signals. *IEEE journal of biomedical and health informatics, 19*(1), 275-281.

Yin, Z., Sulieman, L. M., & Malin, B. A. (2019). A systematic literature review of machine learning in online personal health data. *Journal of the American Medical Informatics Association, 26*(6), 561-576.

Zhu, X., & Goldberg, A. B. (2009). Introduction to semi-supervised learning. *Synthesis lectures on artificial intelligence and machine learning, 3*(1), 1-130.